%% file: paper.tex
\documentclass{article}

\usepackage{arxiv}

\usepackage[utf8]{inputenc} 
\usepackage[T1]{fontenc}    
\usepackage{hyperref}       
\usepackage{url}            
\usepackage{booktabs}       
\usepackage{amsfonts}       
\usepackage{nicefrac}       
\usepackage{microtype}      
\usepackage{lipsum}
\usepackage{dsfont}
\usepackage{mathtools}
\usepackage{enumitem}
\usepackage{comment}
\usepackage{graphicx}
\usepackage{amssymb}
\usepackage{natbib}
\usepackage{defs}
\usepackage{subcaption}

\newcommand{\typei}{Type I}

\newcommand{\exampletag}[1]{\texttt{\##1}}

\newcommand{\kernelband}[2]{e^{-#1\left(#2\right)}}

\DeclarePairedDelimiter\floor{\lfloor}{\rfloor}

\graphicspath{{figures/}}

\title{Detecting Social Influence in Event Cascades by Comparing Discriminative Rankers}

\author{
  Sandeep Soni \\
  School of Interactive Computing\\
  Georgia Institute of Technology\\
  Atlanta, GA 30308\\
  \texttt{sandeepsoni@gatech.edu} \\
  \And
  Shaun Ling Ramirez \\
  Insight Data Science\\
  Seattle, WA 98104 \\
  \texttt{veryslr@gmail.com} \\
  \AND
  Jacob Eisenstein\\
  School of Interactive Computing\\
  Georgia Institute of Technology\\
  Atlanta, GA 30308 \\
  \texttt{jacobe@gmail.com} \\
}

\begin{document}
\maketitle

\input{abstract}


\input{introduction}
\input{related}
\input{discriminative}

\input{synthetic}
\input{application}
\input{discussion}
\input{acknowledgments}
\bibliographystyle{plainnat}  
\bibliography{output}


\end{document}

%% file: abstract.tex
\begin{abstract}
The global dynamics of event cascades are often governed by the local dynamics of peer influence.
However, detecting social influence from observational data is challenging due to confounds like homophily and practical issues like missing data.
We propose a simple discriminative method to detect influence from observational data.
The core of the approach is to train a ranking algorithm to predict the source of the next event in a cascade, and compare its out-of-sample accuracy against a competitive baseline which lacks access to features corresponding to social influence.
We analyze synthetically generated data to show that this method correctly identifies influence in the presence of confounds, and is robust to both missing data and misspecification --- unlike well-known alternatives.
We apply the method to two real-world datasets: (1) the co-sponsorship of legislation in the U.S. House of Representatives on a social network of shared campaign donors; (2) rumors about the Higgs boson discovery on a follower network of $10^5$ Twitter accounts. Our model identifies the role of social influence in these scenarios and uses it to make more accurate predictions about the future trajectory of cascades.
\end{abstract}

%% file: introduction.tex
\section{Introduction}
\label{sec:introduction}
The spread of online social behaviors can often be attributed to \emph{peer influence}, in which the adoption likelihood of a behavior increases through exposures of past adoptions by friends in a network~\citep{crane1999diffusion,goel2016social,karsai2014complex,yang2010modeling}.
In experimental settings, influence can be measured by randomly assigning some individuals to be exposed to the adoption decisions of their friends, or by blocking some individuals from viewing such exposures~\citep{bond201261, salganik2006experimental,muchnik2013social}. However, there are many contexts in which experimental methods cannot be applied: for example, it is not possible to measure the effect of peer influence on the passage of real legislation by a randomized experiment. In such cases, we face the more challenging task of detecting peer influence from observational data. This is difficult because influence is confounded with homophily --- the tendency of individuals with similar properties to establish social network connections. Even when peer influence is absent, the presence of homophily can give rise to network-correlated behaviors which are indistinguishable from the effects of social influence~\citep{shalizi2011homophily}. This concern can be partially addressed by techniques from causal inference, which create comparisons that approximate a randomized experiment, under various limiting assumptions~\citep[e.g.,][]{anagnostopoulos2008influence,aral2009influence,la2010randomization}. 

However, a key question is whether existing methods are robust to phenomena such as missing data and misspecification. In realistic scenarios ranging from social media analysis to criminology, researchers lack complete records of event cascades: social media platforms may provide only a small sample, or relevant data may go unreported. Second, parametric models such as the Hawkes Process require specifying hyperparameters, such as the timescale of the cascade. In real scenarios, there may be only a single cascade, making cross-validation impossible. As we show, such methods perform poorly when such hyperparameters are incorrectly specified.

To address these issues, we propose a simple Granger-style test for social influence, which builds on \emph{predictive} analysis of social event cascades.
The core of our approach is an online, discriminatively trained ranker, which learns to assign a high rank to nodes that are likely to host the next event in a cascade. This is a simpler problem than modeling the probability of a cascade as a series of time-stamped events. This simplicity enables the application of powerful methods from supervised machine learning, which can account for confounds such as homophily through the inclusion of social network node embeddings in the prediction model~\citep{tang2015line}. To detect social influence, we train two ranking models: $m_0$, the most accurate ranker that we can build without including features that measure social influence (including all known confounds), and $m_1$, which contains all features in $m_0$, plus features that measure social influence. We then compare $m_1$ and $m_0$ on held out data: if $m_1$ is more accurate, then this is evidence for social influence, under the usual assumption of no unobserved confounds. 

We validate this approach on a battery of synthetic data experiments, showing that it is well-calibrated (\typei{} error rate at or below the desired $p$-value) when there is homophily, self-excitation, missing data, and model misspecification. The method of ranker comparison obtains higher statistical power than a classical shuffle test across all scenarios, and outperforms a Hawkes Process goodness-of-fit test under conditions of missing data and misspecification. In addition, the \typei{} error rate of the Hawkes Process test approaches one in some settings involving missing data or misspecification of the temporal kernel.


Our contributions are summarized as follows. First, we propose a simple discriminative method to detect peer influence from observational data (\S\ref{sec:discriminative}), which avoids parametric assumptions about the generative process. Second, we demonstrate the robustness of our approach empirically through several synthetic data experiments under conditions of practical importance (\S\ref{sec:evaluation}). Finally, we show the application of our method on two real-world datasets: the cosponsorship of legislation in the United States House of Representatives, and the spread of ``scientific rumors'' about the discovery of the Higgs boson. The latter dataset demonstrates the ability of our method to detect social influence and predict cascade trajectories in networks with $10^5$ nodes and $10^7$ edges.


%% file: related.tex
\section{Related Work}
\label{sec:related}
Social influence induces a systematic temporal pattern in the order of adoptions.
This insight is leveraged by \cite{anagnostopoulos2008influence}, who propose the \emph{shuffle test} to detect social influence.
In this test, social correlation is measured from the observed sequence of adoptions, and compared to a distribution of social correlation after repeatedly shuffling the order of adoptions.
Shuffling eliminates any systematic patterns present due to social influence, while preserving patterns that are due to homophily.
If the observed social correlation significantly differs from the correlation after shuffling, then the result of the test is compatible with social influence. 
The shuffle test assumes a static underlying network, but it can be generalized to dynamic networks~\citep{la2010randomization}.
The shuffle test offers a simple method to detect social influence, without assumptions about the data generation process.
However, it makes no attempt to incorporate fine-grained temporal information or node-level covariates. 
Our method incorporates these features, and is found to be a more powerful statistical test on a range of synthetic data scenarios.

An alternative to shuffling is proposed by \cite{aral2009influence}, who divide nodes in a network into control and treatment groups, depending on whether they have an adopter friend in their ego network.
To account for homophily, units from both groups are matched using propensity scores based on demographic covariates.
The average adoption rate in both groups is then compared across groups.
As we show in \S{~\ref{sec:discriminative}}, our discriminative learning model can capture homophily in the network structure.
Our method can also incorporate personal attributes, but otherwise uses structural similarity as a proxy for homophily.
Unlike the method of \citeauthor{aral2009influence}, our method is capable of predicting the future trajectory of event cascades.

The principle of Granger causality, which relates causation to improvement in predictive accuracy, has been leveraged to detect influence in observational data~\citep{chikhaoui2015new}. This principle motivates the use of \emph{transfer entropy}, an information-theoretic measure of the reduction in conditional entropy in a user's action by incorporating past actions of its network neighbors~\citep{ver2012information, ver2013information, he2013identifying}. Although effective in detecting influence, these methods do not learn to make predictions and their robustness on conditions of missing data is unknown. Our proposed method is also based on Granger causality, but we show empirically that its robust under practical limiting conditions of missing data.

An alternative predictive approach is to estimate a parametric influence network by modeling one or more timestamped event cascades as a \emph{Hawkes Process}~\citep[HP;][]{hawkes1971spectra,du2013scalable,yang2013mixture,zhao2015seismic}.
If the social excitation parameters are estimated to be non-zero, this is consistent with social influence~\citep{xu2016learning}; alternatively, a goodness-of-fit test can be used to compare nested parametric Hawkes Process models, with and without access to features of the social network~\citep{goel2016social}.
Our discriminative learning model is similar in some respects: it is a learning based model, and also builds on the assumption that past events modulate the rate of future events.
However, the Hawkes process is a \emph{generative} model of events in a cascade, while our proposed approach is discriminative.
As we show, violations of the assumptions of the HP generative model lead to incorrect inferences, yielding high \typei{} error rates.
In contrast, our discriminative modeling approach is agnostic to the event generation process, and is robust to missing data and relatively insensitive to misspecification.
While there are targeted solutions for misspecification and missing data~\citep[e.g.,][]{he2016influence, duong2011modeling, lokhov2016reconstructing, xu2017learning}, there is no general approach that makes the Hawkes process immune to all such concerns. 

Granger causality can be applied to Hawkes process models to learn pairwise influence parameters~\citep{xu2016learning}. Our approach is aimed at testing the presence of social influence at the network scale, rather than reliably estimating pairwise influence.
This makes it possible to formulate our approach through the relatively simpler framework of model comparison, rather than attempting to induce group-sparse excitation parameters.   
Other predictive models focus on properties such as the cascade's ultimate size~\citep[e.g.,][]{cheng2014can} or the susceptibility of individuals~\citep[e.g.,][]{du2013scalable}, but these models do not easily lend themselves to the detection of social influence as a causal phenomenon.

The Granger causality principle is also the basis of some proposed discriminative models that make predictions about user actions from the past actions of their neighbors~\citep{tang2013confluence, zhang2015influenced, qiu2018deepinf}. However, these models pre-suppose the existence of social influence to make predictions in contrast to our objective of identifying social influence in the presence of confounds.


%% file: discriminative.tex
\section{Discriminative Model}
\label{sec:discriminative}

Our objective is to detect social influence given a cascade of time\-stamped events from individual nodes in a social network.
We do this by solving an auxiliary task: predicting which node will be the next to be activated in the cascade.
Specifically, we learn a discriminative ranking function, which scores each node by weights on features of the individual nodes and the cascade history.
Observed confounds --- variables that predict both the presence of social network connections and participation in the event cascade --- can be incorporated into the ranker as features on individual nodes.
(The usual assumption of no unobserved confounds is still required~\citep{shalizi2011homophily}.)
The question of whether there is social influence is then transformed to the proxy question of whether the prediction task is aided by the inclusion of features that measure social influence.
This is similar to the principle of \emph{Granger causality}~\citep{granger1969investigating}, which states that $X$ Granger-causes $Y$ if prediction of $Y$ is aided by knowledge of $X$.\footnote{Note that ranking the nodes by their likelihood of being activated next in a sequence is different from ranking nodes by their influence in the network~\citep{chen2012identifying}.}


\subsection{Features for discriminative ranking}
\label{sec:discriminative-scoring}
To order all nodes in the network by their likelihood of being the next node to be activated in the cascade, we compute a scoring function which depends on static features of the node, and the history of the cascade. The static features can account for confounders, which relate to the base propensity of each node to participate in the cascade; the history features can account for both self-excitation and social influence. The score for node $i$ at time $t$ is:
\begin{align}
\label{eq:ranker-scoring-function}
\sco_{i}\left(t; \thetabf, \phibf \right) &= h\left(\mathbf{g}_{i};\phibf\right) + \sum\limits_{e:t_e < t} \thetabf \cdot \mathbf{f}_{\edge{s_e}{i}} \, \kappa(t-t_e; \omega) ,
\end{align}
where both $\mathbf{g}_i$ and $\mathbf{f}_{\edge{\boldsymbol{\cdot}}{i}}$ are node-specific feature vectors for node $i$, and $e$ indicates an event with source $s_e$ and time $t_e$. 

The features $\mathbf{g}$ capture intrinsic properties of the node; these features are static in all subsequent evaluations, but the model generalizes trivially to dynamic features. The function $h$ transforms the features into a scalar value, and is parameterized by $\phibf$; this function could be, e.g., a simple inner product of features and weights, or a multilayer neural network. The features $\mathbf{f}$ capture the properties of each dyad. For the ordered dyad $\edge{s_e}{i}$, the features are written $\mathbf{f}_{\edge{s_e}{i}}$, and the associated weights are written $\thetabf$. As in the Hawkes Process, these features are weighted by a temporal kernel $\kappa$, with bandwidth parameter $\omega$.

The ranker is applied in an online fashion, recomputing scores for each node at the time of every new event. While the scoring function is closely related to the intensity function of the Hawkes Process (see~\autoref{eq:mhp-cond-int}) there are two key differences. First, the intensity of a generative model like a Hawkes Process has to be non-negative at any time, however, the scoring function of our ranker is free from such constraints and can yield negative scores. Second, in addition to calculating the probability of an event, the Hawkes Process has to further calculate the \emph{survival} probability resulting in the computation of complicated integrals of the intensity function over inter-event periods. In contrast, our ranker is online and we are concerned only with the order of the nodes for the next event. The ranker therefore avoids integrating the intensity function over inter-event periods, simplifying estimation.

\subsubsection{Node-level features.}
The features $\mathbf{g}$ can act as a proxy for homophily between the nodes, which might otherwise confound the detection of social influence.
In some cases, specific covariates are available: for example, the political party of each legislator, or the age of members of a social network.
But in other cases, the relevant covariates are unknown.
In this case, we assume that the latent features that drive the event cascades are related to the properties that drive the formation of social network ties; it is just these features that risk confounding the estimation of social influence.
Network structural properties can be captured by  computing \emph{node embeddings}, where neighbors in the network are nearby in the embedding space.
There are several ways of calculating the node embeddings~\citep{grover2016node2vec,tang2015line}; in the analyses that follow, we use spectral embeddings obtained from the graph Laplacian~\citep{donetti2004detecting}.
To increase the expressive power of the ranker, each node embedding is transformed into a scalar by a feedforward neural network, which finds the parts of the network that are most susceptible to participating in the cascade. During learning, the node embeddings are finetuned by minimizing the ranking loss objective described in \S{\ref{sec:discriminative-learning}}.

\subsubsection{Dyadic features.} 
In addition to the intrinsic node features, the ranker also utilizes dyadic features extracted from past events.
As in the Hawkes Process, these features are multiplied by a decay kernel $\kappa$, and aggregated over the entire history of past events.
The decay kernel captures the intuition that events in the distant past should have less impact on the score than more recent events.
In our experiments, we use a simple exponential decay kernel $\kappa(t) \propto \exp(-\omega t)$, with $\omega$ acting as a bandwidth parameter.
However, the decay kernel can be a more complex function, such as a linear combination of simple kernels, or a Gaussian process~\citep{bacry2012non}.
We incorporate two dyadic features: self-excitation, which is the tendency for a node to repeatedly activate after the first activation; and social influence, which is the tendency for a node to be activated by exposure from its neighbors.  The complete set of features used in our model and their description is given in Table~\ref{table:features}.

\subsection{Learning}
\label{sec:discriminative-learning}
Ranking weights can be estimated by minimizing the a pairwise loss called WARP~\citep[Weighted Approximate Rank Pairwise;][]{usunier2009ranking}. To avoid evaluating the quadratic number of node comparisons, we apply the well-known WSABIE approximation~\citep{weston2011wsabie}, which samples pairs until a violation is found. 

Let $\Acal(t)$ be the set of nodes with an event at time $t$, and let $\rank_i(t)$ be the rank of node $i$ according to the scoring function $\sco$. We define $\rank^{1}_i(t)$ as the \emph{margin-penalized rank},
\begin{equation}
\label{eq:rank}
  \rank^{1}_i(t) = \sum_{j \notin \Acal(t)} \mathds{1}[1 + \sco_j(t) > \sco_i(t)],
\end{equation}
which is equal to the number of margin violations for active node $i$, where $\mathds{1}[\cdot]$ is the indicator function. Using these terms, we can compute the WARP loss at time $t$:
\begin{align}
  \label{eq:WARP-loss}
  \sum_{i \in \Acal(t)} \frac{L(\rank^{1}_i(t))}{\rank^{1}_i(t)} \sum_{j \notin \Acal(t)} (1 - \sco_i(t) + \sco_j(t))_{+},
\end{align}
where $(\cdot)_{+}$ is shorthand for $max(\cdot, 0)$ and $\loss(k)$ is a ranking error function. We choose a simple ``one-best'' error function, whose value is one if $k>1$ and zero otherwise; however, the formalism enables a flexible class of alternative error functions~\citep{usunier2009ranking}. 

Calculating the loss in (\ref{eq:WARP-loss}) naively is inefficient for a large network. The WSABIE algorithm approximates this loss by repeatedly sampling inactive nodes randomly until a node is found which violates the margin constraint. The rank of the activated node is approximated by the inverse of the number of samples required to find such a node. The overall loss function used by the WSABIE algorithm is given by, 
\begin{align}
\label{eq:ranker-loss-function}
\sum_{i \in \Acal(t)} \loss\left(\floor*{\frac{M-1}{\text{trials}_{i,t}}}\right) \times \,( 1 - \sco_{i}(t) + \sco_{j}(t))_+,
\end{align}
where $M$ is the number of nodes, $\text{trials}_{i,t}$ is the number of trials required to find a margin violation, and $j$ is the violating node.
This loss is calculated at each time, and the scoring function parameters are updated by taking a gradient step to minimize the loss.




\begin{table}
\caption{The complete set of features used by the ranker. Note that the dyadic features are scaled by the decay kernel and are aggregated for the entire history.}
  \label{table:features}
  \centering
  \begin{tabular}{cccp{10cm}}
    \toprule
    Feature &Type & Range & Description\\
    \midrule
    \textit{emd} & Node & $\mathbb{R}^K$ & The spectral embedding for each node $i$ \\
    \textit{self} & Dyad & $\{0,1\}$ & Self excitation; active for node $i$ if $j$ has a previous event and $j=i$\\
    \textit{social} & Dyad & $\{0,1\}$ & Social feature; active for node $i$ if $j$ has a previous event and $A_{\edge{j}{i}}=1$ \\
  \bottomrule
\end{tabular}
\end{table}

\subsection{Model Comparison}
\label{sec:model-comparison}
To test for social influence, we compare the performance of alternative ranking models $m_0$ and $m_1$, which are identical in all respects except one: $m_1$ includes features that are activated under the condition of social influence, and $m_0$ does not. In particular, $m_1$ includes a feature that fires for node $i$ at time $t$ if any of the social network neighbors of $i$ have an event at time $t' < t$. This feature is included in the vector $\mathbf{f}_{\cdot \to i}$ in \autoref{eq:ranker-scoring-function}. Each ranker is then applied to heldout data, and evaluated according to a ranking metric. In most cases, we use mean reciprocal rank (e.g. all cases in the synthetic data experiments).

To determine whether the performance difference between $m_1$ and $m_0$ is unlikely to have arisen due to chance alone, we must apply a statistical significance test to compare their performance on a heldout test set.
Of the various statistical significance tests that have been proposed for rankers~\citep[e.g.,][]{cormack2006statistical,sakai2006evaluating,wilbur1994non}, we select the non-parametric permutation test~\citep{smucker2007comparison}.
In this test, the predictions between the two rankers are repeatedly exchanged (permuted), to create an empirical distribution of the difference in ranker performance under the null hypothesis that the two rankers are identical. 
The right-tailed $p$-value is the fraction of permutations in which the difference in ranker performance was greater than the observed difference in the original unpermuted data.
This tests the null hypothesis that adding social influence features does not improve ranking accuracy on heldout data, which is the Granger-based proxy of the null hypothesis of no social influence. 
In the following section, we demonstrate the reliability of this proxy through a series of experiments on synthetic data.

%% file: synthetic.tex
\section{Synthetic Data Evaluation}
\label{sec:evaluation}

To determine the validity and efficacy of our proposed test for social influence, we evaluate it on a set of synthetic cascades generated over a real social network. Some cascades are generated without social influence, to test the Type I error rate of our method (incorrect rejection of the null hypothesis); other cascades are generated with social influence, to test the power (correct rejection of the null hypothesis). In both cases, we consider the impact of homophily, self-excitation, missing data, and model misspecification. The use of synthetic data makes it possible to quantify these characteristics precisely, before moving to real data in the next section. 
\subsection{Data}
\label{sec:generative}

Event cascades are generated using a multivariate Hawkes Process (HP), which is an inhomogenous Poisson process in which the intensity is modulated by the past history of events~\citep{hawkes1971spectra}.
The intensity of node $i$ at time $t$ represents the instantaneous activation rate, and is given by,
\begin{align}
\label{eq:mhp-cond-int}
\lambda_{i}\left(t\right) = \mu_{i} + \sum\limits_{j=1}^{M}\sum\limits_{\substack{\tau: \tau \in \Tcal_{j}}} \alpha_{\edge{j}{i}}\kernelband{\omega}{t-\tau}\mathds{1}[\tau < t],
\end{align}
where $\mubf$, $\alphabf$, and $\omega$ are the parameters of the intensity function and $\Tcal_{j}$ is the set of events from node $j$. 
The base generation rate is $\mubf$, and $\alphabf$ is the matrix of pairwise excitation parameters between nodes.

To generate cascades under conditions of social influence and various confounds, we specify these parameters as follows:
\begin{align}
\label{eq:influence-generative}
\alpha_{\edge{j}{i}} &= a\mathds{1}[j=i] + bA_{\edge{j}{i}}\mathds{1}[j\neq i] \\
\label{eq:homophily-generative}
\mu_i &= \sigma\left(\beta v_i^{(2)} + \eta\right)
\end{align}
where $v^{(2)}$ is the second eigenvector of the Laplacian matrix of the network; $\sigma(\cdot)$ is the sigmoid function, which ensures that the base rate is positive; and $A$ is the adjacency matrix of the network. We can then generate cascades under various conditions of interest:

\begin{description}[leftmargin=*]
  \item[Social influence.] By setting $b=0$, we generate cascades without social influence. These cascades are used to measure the Type I error rate of our test. As $b$ increases, so does the impact of social influence. Throughout the evaluation, $b$ is varied between $0$ to $1$ in increments of $.1$, depending on the experimental setting.
  \item[Homophily.] The similarity of nodes $(i,j)$ is captured by the similarity between the components of the second eigenvectors $v_i^{(2)}$ and $v_j^{(2)}$. By conditioning the base adoption rate of these nodes on this parameter, it is possible to generate cascades in which events are strongly correlated with the network, even without social influence. This corresponds to the case in which nodes $i$ and $j$ form a friendship because they both share an interest, and then participate in cascades that reflect that same interest. By varying the parameter $\beta$, the importance of homophily in shaping the cascades can be increased or decreased. $\beta$ is varied between $0$ to $7$ in increments of $1$, depending on the experiment condition. In all experiment settings, $\eta$ is set to $-5$ for all the nodes.
  \item[Self excitation.] For some types of cascades, nodes can participate repeatedly. Self excitation occurs when a node's own participation in the cascade spurs further participation in the future. For instance, after learning a new slang or a hashtag and using it once, a user is likely to repeat its usage~\citep{goel2016social}. This tendency can be controlled by setting $a>0$. We vary $a$ between $0$ to $1$ in increments of $.25$, depending on the experiment condition.
  \item[Network structure.] Rather than generating a synthetic network, we use a real ``mention'' network from Twitter. This static and directed network was constructed as follows. First, we select all individuals who used a partisan political hashtag (e.g., \exampletag{clintonkaine2016}) between October 1-15, 2016. Then we identify all individuals who were mentioned by someone in this initial set. The directed edge $i \rightarrow j$ indicates that $i$ mentioned $j$ on Twitter during this time period. We then select the largest weakly connected component as the underlying network for all synthetic cascades.
\end{description}

\subsection{Evaluation Metrics}
\label{sec:sythetic-data-evaluation-metrics}
We evaluate the ranking test on two metrics: \textit{validity} and \textit{power}.

\subsubsection{Validity}
A test is statistically valid if its $p$-values are well calibrated: at the threshold $p = \alpha$, the test should reject a true null hypothesis with probability less than or equal to $\alpha$. To establish validity, we evaluate the test on cascades where the null hypothesis is known to be true. The rate at which the test rejects the null hypothesis is the Type I error rate. 

\subsubsection{Power} 
A test has high statistical power if it consistently rejects a false null hypothesis.
Failure to reject a false null hypothesis leads to a Type II error and the power is the probability of not making a Type II error.
To establish power, we evaluate the test on multiple cascades where the null hypothesis is known to be false.
The rate at which the test fails to reject the null hypothesis is the Type II error rate.
An ideal test should be valid under all conditions and should have high statistical power. 

\subsection{Methods}
\label{sec:synthetic-data-baselines}
We compare the performance of the ranking test against two well known alternatives: (1) the shuffle test from Anagnostopoulos \etal~\citep{anagnostopoulos2008influence}, and (2) a test that compares the goodness of fit between two parametric Hawkes Processes (HP): one with access to social influence parameters, and one in which these parameters are clamped to zero. We now describe these tests in detail.

\subsubsection{Shuffle test}
The shuffle test infers social influence by calculating a measure of social correlation from an observed cascade, and comparing it with the distribution of the same measure after repeatedly shuffling the order of events in the cascade.
This shuffling breaks the effect of social influence, providing an estimate of the amount of social correlation that can be attributed to factors other than influence.
We use the \emph{infection risk} as our measure of social correlation, which is calculated as the ratio of \emph{adopters} to \emph{innovators}:
\begin{itemize}
\item adopters are nodes that are activated only after at least one of their network neighbors were activated;
\item innovators are nodes that are activated before any of their network neighbors.
\end{itemize}
To calculate the infection risk, only the first activation of a node is considered.
The infection risk has been used as a measure for social correlation in other studies~\citep[e.g.,][]{aral2009influence}.
As there is no learning involved, the computational expense for this test is not huge. The statistics involved in the calculation of the infection risk can be computed efficiently using vectorized operations.

\subsubsection{HP goodness-of-fit test}
The Hawkes Process (HP) can be used to test for social influence by comparing the goodness-of-fit between two nested models: an HP that includes social influence parameters, and an HP that does not~\citep[e.g.,][]{goel2016social}. To perform this test, we estimate the parameters of the HP described in Equations~\ref{eq:mhp-cond-int}, \ref{eq:influence-generative} and \ref{eq:homophily-generative} from training data.
To test whether the goodness-of-fit improved significantly by the addition of the \textit{social} feature, we use the likelihood ratio test~\citep{wilks1938large}.
Since the synthetic data was truly generated from a Hawkes Process, we expect this test to perform well, unless there is missing data or misspecification in the data generation process. 

\input{experiments}

%% file: experiments.tex
\subsection{Results}
\label{sec:experiments}

\subsubsection{Full data}
We first consider the case when every test has access to full data: all the events in the cascade and all the edges in the network are known to each test.
We also assume that the generative process is correctly specified, meaning that the exponential decay kernel and the bandwidth parameter that modulate the rate of generation of future events are known during learning and testing. We will relax these assumptions later.

\begin{figure}
    \begin{subfigure}[t]{0.49\textwidth}
        \centering
        \includegraphics[width=\textwidth]{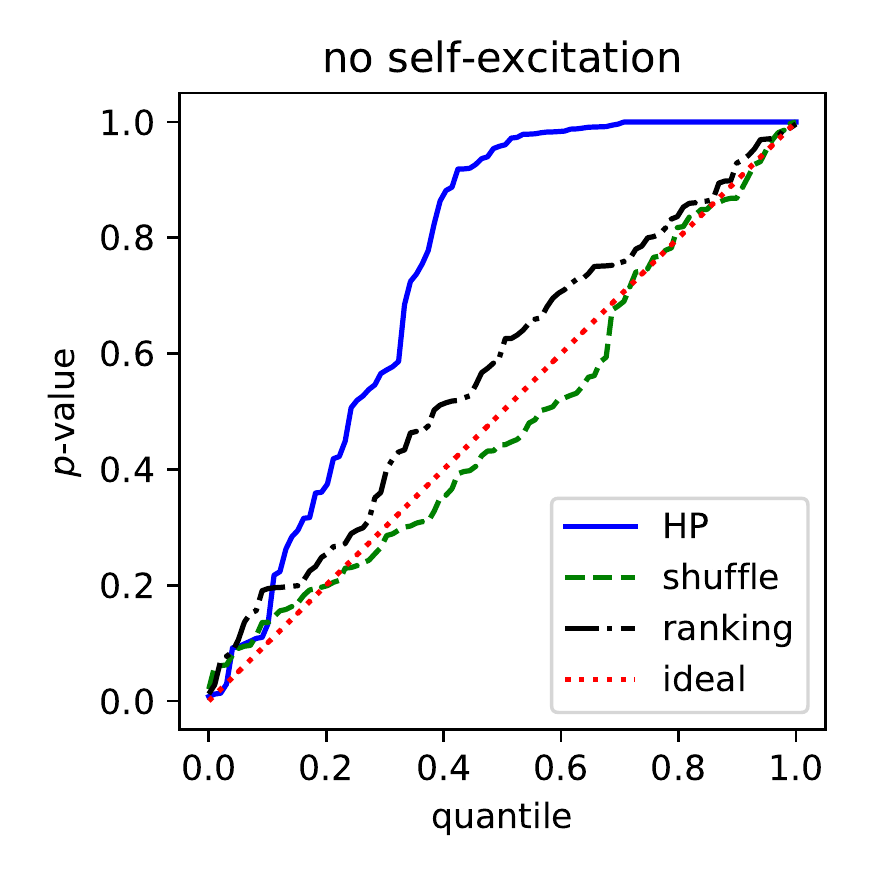}
        \label{fig:qq-noself}
    \end{subfigure}%
    \begin{subfigure}[t]{0.49\textwidth}
        \centering
        \includegraphics[width=\textwidth]{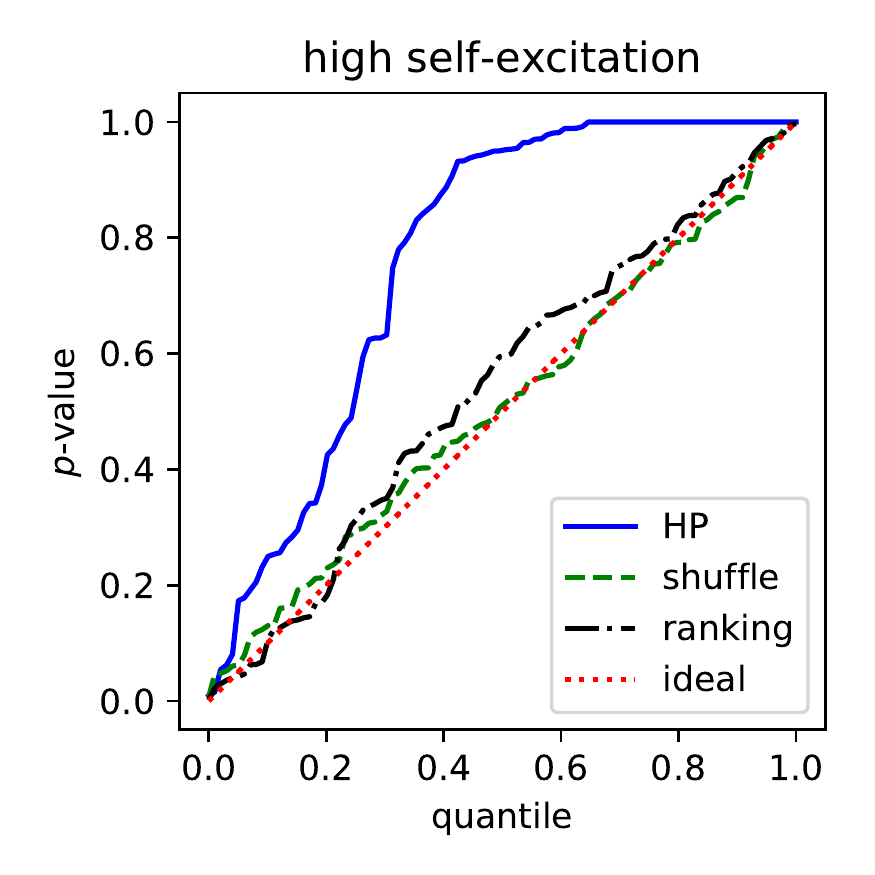}
        \label{fig:qq-highself}
    \end{subfigure}
    \vspace*{-0.8cm}
    \caption{Quantile-Quantile plots of $p$-values for all the three tests under high homophily. The dotted red line shows the expected plot for an ideal well-calibrated test.}
    \label{fig:type1-fullmodel}
\end{figure}

\paragraph{Validity.}
To check for validity, 100 cascades of 5000 events each are generated under conditions of no social influence ($b=0$), varying homophily ($\beta >0$) and varying self-excitation ($a>0$).
The null hypothesis of no social influence is true by design for every such cascade.
\autoref{fig:type1-fullmodel} shows the calibration of the tests when there is high homophily ($\beta=7$) and low (left) or high self-excitation (right).
Both the ranking and shuffle tests are well-calibrated.
The goodness-of-fit test of HP produces conservative $p$-values, but satisfies the condition of validity, which is that the Type I error rate is bounded by $p$-value.
All three tests have low Type I error rates even under different amounts of homophily and self-excitation.

\begin{figure}
    \begin{subfigure}[t]{0.49\linewidth}
        \centering
        \includegraphics[width=\textwidth]{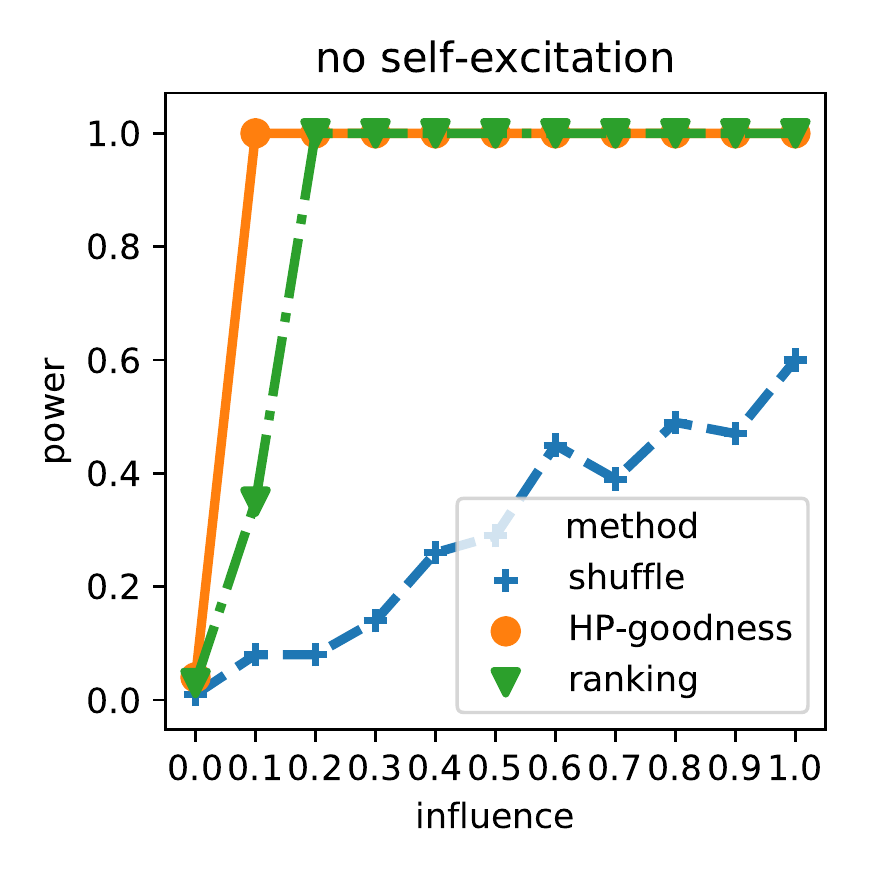}
        \label{fig:type1-mediumself}
    \end{subfigure}
    ~
    \begin{subfigure}[t]{0.49\linewidth}
        \centering
        \includegraphics[width=\textwidth]{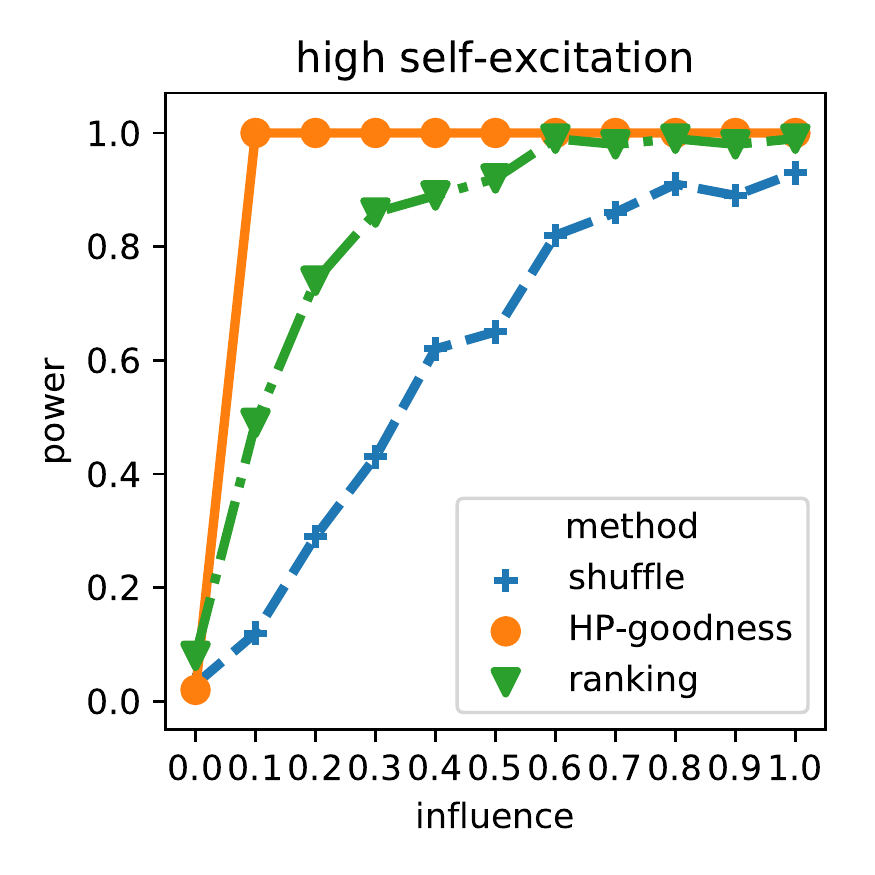}
        \label{fig:type1-highself}
    \end{subfigure}
    \vspace*{-0.8cm}
    \caption{Power for all the three tests under high homophily.}
    \label{fig:type2-fullmodel}    
\end{figure}

\paragraph{Power.} To check for statistical power, 100 cascades of 5000 events each are generated by varying social influence at fixed values of homophily and self-excitation.
As shown in \autoref{fig:type2-fullmodel}, the power increases with social influence for all tests, as expected.
The \hptest{} test is the most powerful across these conditions, and the \shuftest{} test is least powerful. 
As noted above, since the HP captures the true generative process, it should outperform the shuffle test, which is agnostic to the generative process.
Because the ranking test also outperforms the \shuftest{} test, in the subsequent evaluations we only compare the ranking test to \hptest.

\subsubsection{Model misspecification}
\begin{figure}
    \begin{subfigure}[t]{0.49\linewidth}
        \centering
        \includegraphics[width=\textwidth]{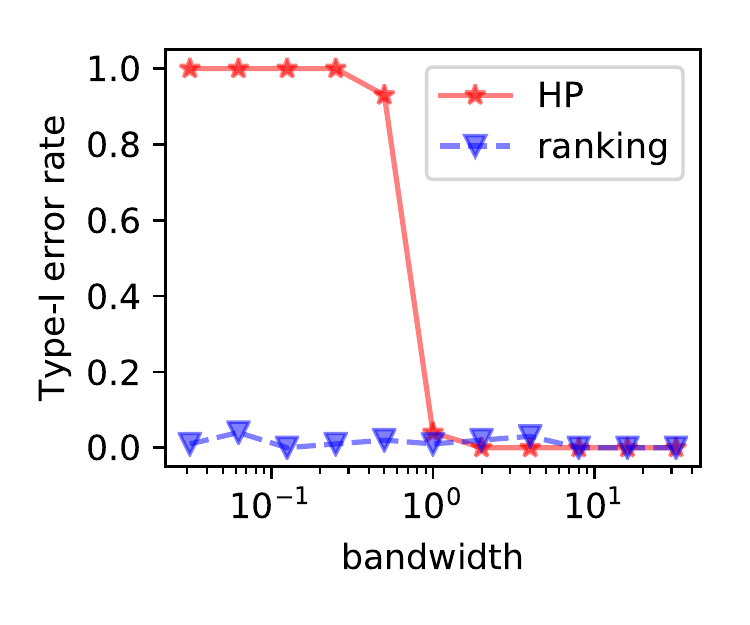}
        \label{fig:misspec-typeI}
    \end{subfigure}
    ~
    \begin{subfigure}[t]{0.49\linewidth}
        \centering
        \includegraphics[width=\textwidth]{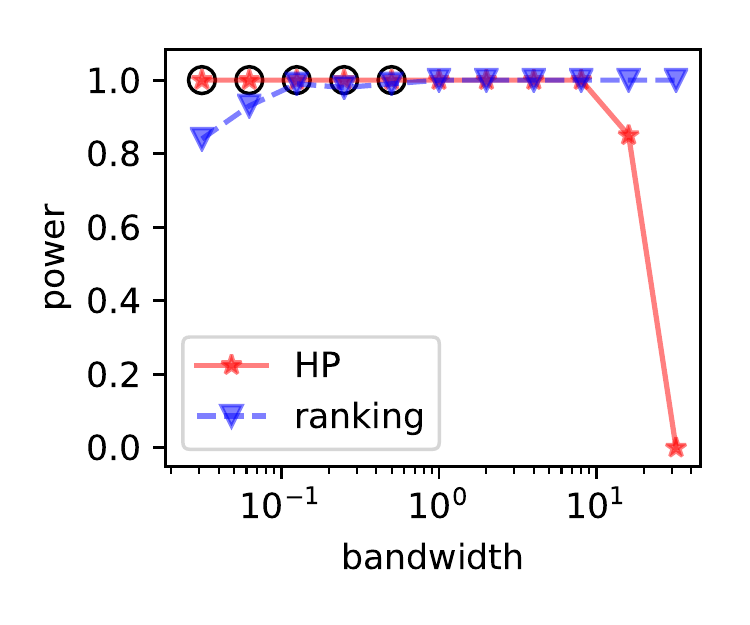}
        \label{fig:misspec-power}
    \end{subfigure}
    \vspace*{-0.8cm}
    \caption{Type I error rate (left) and power (right) for HP and ranker test under misspecification of kernel bandwidth($\omega$). The HP test is invalid when $\omega < 1$; this is indicated by black circles around relevant data points when showing power.}
    \label{fig:model-misspecification} 
\end{figure}

Next we examine how the tests fare under misspecification, focusing on the bandwidth parameter ($\omega$) for the temporal decay kernel.
We generated 100 cascades of 5000 events each for the influence ($b=1$) and no-influence condition ($b=0$), both with highest homophily and self-excitation ($a=0.5$; $\beta=7$).
During generation, the bandwidth was fixed ($\omega^*=1$), but we assumed that both the HP learner and the ranker did not know this true value of the bandwidth parameter.
We then varied the values of $\omega$ used by these tests, to understand their effect on validity and the statistical power of both the tests.
Varying the bandwidth parameter has a natural interpretation: as $\omega$ decreases, the scope of the history is effectively widened; increasing this parameter has the opposite effect. 
The results for both validity and power are shown in \autoref{fig:model-misspecification}.

Increasing the bandwidth parameter causes the temporal kernel to decay sharply, nullifying the effect of spurious activations of neighbors from the past, limiting Type I error but also power.
The Type I error rate decreases with bandwidth for exactly the opposite reason:
as the bandwidth parameter decreases, the impact of past events is larger on each node's activation. Irrelevant activations of neighbors in the distant past may be mistakenly deemed consequential, leading the test to overestimate the role of social influence.
The \hptest{} model is quite sensitive to this parameter, and misspecification severely undermines the validity of the test for practical purposes.

In contrast, the ranking test is robust to misspecification: for a wide range of bandwidth values, the ranker maintains a low Type I error rate and has considerable power.
This is because the ranking objective requires only that the relative order of each node is maintained.
Spurious events in the distant past affect many nodes, and the resulting changes in rank are not significant.
While there is research on augmenting the Hawkes Process with nonparametric kernels that are learned from data~\citep[e.g.,][]{zhou2013learning}, such methods are complex to implement, and require large amounts of training data. For a parametric kernel, the bandwidth can be learned by cross-validation, but only if multiple cascades are available and if the parameter is guaranteed to be static over time. In contrast, the ranking test can easily incorporate multiple kernels into the discriminative ranking function and is completely online, thereby having the flexibility to learn complex triggering patterns from a single cascade.  

\subsubsection{Missing data}
We have thus far assumed that all events in the cascade are available.
There are various reasons that this assumption can be violated in real data: full data collection is often too expensive; there are rate limits on collecting events from public APIs for sites such as Twitter; individuals may erase past events in their history; data may be lost accidentally, as can happen when a server crashes during data acquisition; data collection may have begun only after the cascade was initiated. Incomplete data can diminish statistical power, giving the appearance of unprompted innovations to events that were in fact socially motivated. 
To quantify this phenomenon, we generate cascades with two different types of missing events: events missing at random, and events missing in contiguous blocks.\footnote{Other types of missing data, such as missing edges or missing nodes, should also be considered~\citep{duong2011modeling, linderman2017bayesian}. We leave the validation of our test against these types of missing data for future work.} 


\begin{figure}
    \begin{subfigure}[t]{0.49\linewidth}
        \centering
        \includegraphics[width=\textwidth]{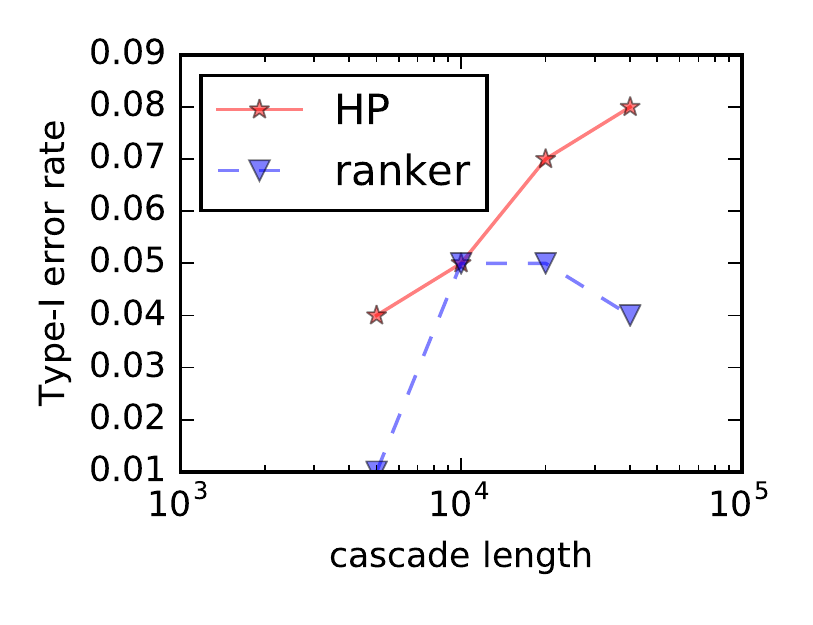}
        \label{fig:randevents-typeI}
    \end{subfigure}
    ~
    \begin{subfigure}[t]{0.49\linewidth}
        \centering
        \includegraphics[width=\textwidth]{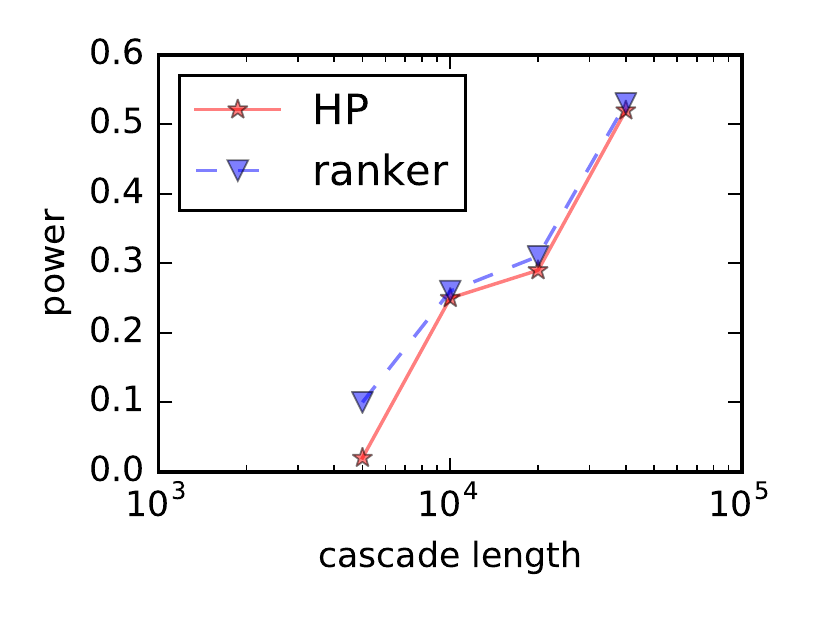}
        \label{fig:randevents-power}
    \end{subfigure}
    \vspace*{-0.8cm}
    \caption{Type I error rate and power for HP and ranker test on 99 \% random missing events on cascades of increasing length. The ranker's power is always higher than HP particularly on cascades of shorter length (5000 events). The Type I error rate for both methods is around 0.05.}
    \label{fig:events-sampling} 
\end{figure}
\subsubsection*{Random missing events}
\label{sec:random-missing-events}
We generate 100 cascades of varying lengths, and randomly drop events in each cascade at a specified rate of 99\% before presenting them to the learners. 
This sampling rate matches a published estimate of the fraction of tweets included in Twitter's streaming API~\citep{morstatter2013sample}.
\autoref{fig:events-sampling} shows the power of the \hptest{} and ranking tests, as a function of cascade length.
While both tests increase in power with the cascade length, the \hptest{} is marginally less powerful across all cascades, particularly on short cascades. 
The Type I error rate for both tests is around 0.05 even as cascade length increases, and is therefore not shown.

\subsubsection*{Doubly censored events}
\label{sec:doubly-censored-events}
\cite{xu2017learning} consider an alternative scenario, in which contiguous blocks of events are missing both at the start and the end of the cascade, resulting in \emph{short doubly censored} sequences.
The censoring scenario is relevant to scenarios like social media analysis; for example, the Twitter API restricts the collection of data from the distant past. 
We generated 100 cascades of different lengths and dropped 99\% of events, with half of the dropped events in a contiguous block at the beginning and the remaining half in a contiguous block at the end of the cascade.
On censored cascades, the goodness-of-fit test for HP suffered from a high Type I error rate, limiting its validity.
This bias towards overestimating social influence is consistent with findings from \cite{xu2017learning}, who note that maximizing the likelihood of an HP can lead to overfitting in this scenario.
Conversely, the validity of the ranking test was not affected by censoring, because the ranking test does not rely on explaining the temporal distribution of events.
However, the loss of information due to censoring does affect the power of the ranking test, especially if the length of the cascade is small; this is a natural consequence of the difficulty of estimation from limited data.

\subsection{Summary}
To summarize the main findings from synthetic data:
\begin{itemize}
\item The shuffle test is underpowered, because it uses only the first-time activations for each node, and because it ignores the time between activations.
\item The \hptest{} test has low Type I error rate and high power when (a) the model is correctly specified and (b) complete data is available.
\item However, the \hptest{} test is highly sensitive to misspecification of the time kernel, and to missing events. These cases can make the test statistically invalid or hurt its power.
\item The ranking test is robust across all scenarios: it is valid in all scenarios, and nearly matches the \hptest{} test's power when complete data is available; model misspecification has little impact on its validity or power; and it is reasonably powerful even under different types of missing data conditions.   
\end{itemize}

%% file: application.tex
\section{Real Data Evaluation}
\label{sec:application}
The synthetic data experiments demonstrate the power and validity of our ranker in detecting social influence. In this section, we highly its \emph{predictive} ability by applying it to two real world datasets: cosponsorship of bills in the U.S. House of Representatives (\S{~\ref{sec:cosponsorship}}) and spread of rumors around the discovery of the Higgs boson particle (\S{~\ref{sec:scientific-rumor}}).
\begin{table}
\caption{The statistics for the network in both the datasets.}
  \label{table:network-statistics}
  \centering
  \begin{tabular}{p{5.5cm}p{3.5cm}p{3cm}}
    \toprule
    Statistic & Legislator Network & Friends-Follower Network\\
    \midrule
    Type & undirected & directed\\
    Total nodes & 436 & 456626\\
    Total edges & 31323 & 14855842\\
    Giant component size (\% nodes) & 98.1 &78.9\\
    Average degree & 143.68 & 39.15\\
  \bottomrule
\end{tabular}
\vspace{-0.2cm}
\end{table}
\subsection{Legislative co-sponsorship and political finance networks}
\label{sec:cosponsorship}
A key step in the legislative process is when a bill receives endorsement from legislators besides its original author, called cosponsors. Cosponsorship decisions are important markers of wider support; they signal the expertise of the original sponsor, and provide information about the bill's content and the party, ideological, or constituency base for whom the bill advocates. However, cosponsorship is also a low-cost means of position taking, reflective of favor-trading, vote-buying, and special interest politics~\citep{kessler1996dynamics}. An open question is whether cosponsorship decisions are influenced by campaign donations, for example by facilitating special access to legislators~\citep{kalla2016campaign}. To test this question, we construct an affiliation network among legislators if they share common campaign donors, and apply our discriminative ranking test to sequences of cosponsorship decisions. The detection of influence on this network would be compatible with the hypothesis that campaign donations influence cosponsorship decisions.

\subsubsection{Cascade data}
We collect cosponsorship sequences on bills introduced in the 115th U.S. House of Representatives, using ProPublica's Congress API.\footnote{https://projects.propublica.org/api-docs/congress-api/}
We only consider bills from the House of Representatives, and ignore resolutions, since these are not presented to the President to become law. 
We filter out bills that have fewer than five or more than 200 cosponsors, resulting in a total of 1022 bills. 
The cosponsorship sequence for each bill -- a sequence of events with a legislator as the source and the date as the time -- is considered as a separate cascade. The typical average number of events per cascade is eight.

\subsubsection{Social network data}
For every representative, we collect a list of their top 20 campaign donors from public sources.\footnote{https://www.opensecrets.org/}
We construct a social network such that a pair of legislators is connected if they share a top donor. 
Some statistics for this network are described in Table~\ref{table:network-statistics}.
For every legislator, we include two node-level covariates: party affiliation and the state they represent. 

\subsubsection{Problem setup and evaluation}
To obtain evidence of network influence, we use the ranking test to compare predictive performance of two rankers. We randomly divide the set of bills into a training and test set of equal sizes. All events from bills in the training set are used for estimating the parameters of the rankers, and both rankers make predictions for every event from bills in the test set. One distinctive feature of this data is the temporal resolution of the cascades: events are timestamped only by date, and there are often multiple new cosponsors on the same date. This creates a problem during training, since our discriminative ranker is based on the mean reciprocal rank (MRR), which assumes a single event at any time.\footnote{It is possible to formulate the WARP loss with an alternative error function, allowing multiple events at each ``query'', or time point. We leave this for future work.} We resolve this by adding a small amount of random ``noise'' to the time of each event during training to separate them but preserve their order of occurrence. We evaluate predictive performance by calculating the mean average precision (MAP), rather than MRR, for each bill in the test set.

\subsubsection{Baseline}
The features from Table~\ref{table:features} are not all applicable to cosponsorship cascades. In particular, self-excitation is not applicable since representatives do not cosponsor the same bill twice; however, cosponsorship may be affected by dyadic features such as party or state affiliation. For the baseline ranker, we use two dyadic features for every node, which are activated if past cosponsors are from the same party or same state respectively.
We also learn per-representative parameters that capture the tendency of each legislator to cosponsor legislation. We then add the social feature from Table~\ref{table:features} to the set of features in the baseline to construct our socially-augmented ranker.

\subsubsection{Results}
\begin{figure}
  \centering
  \includegraphics[width=0.6\textwidth]{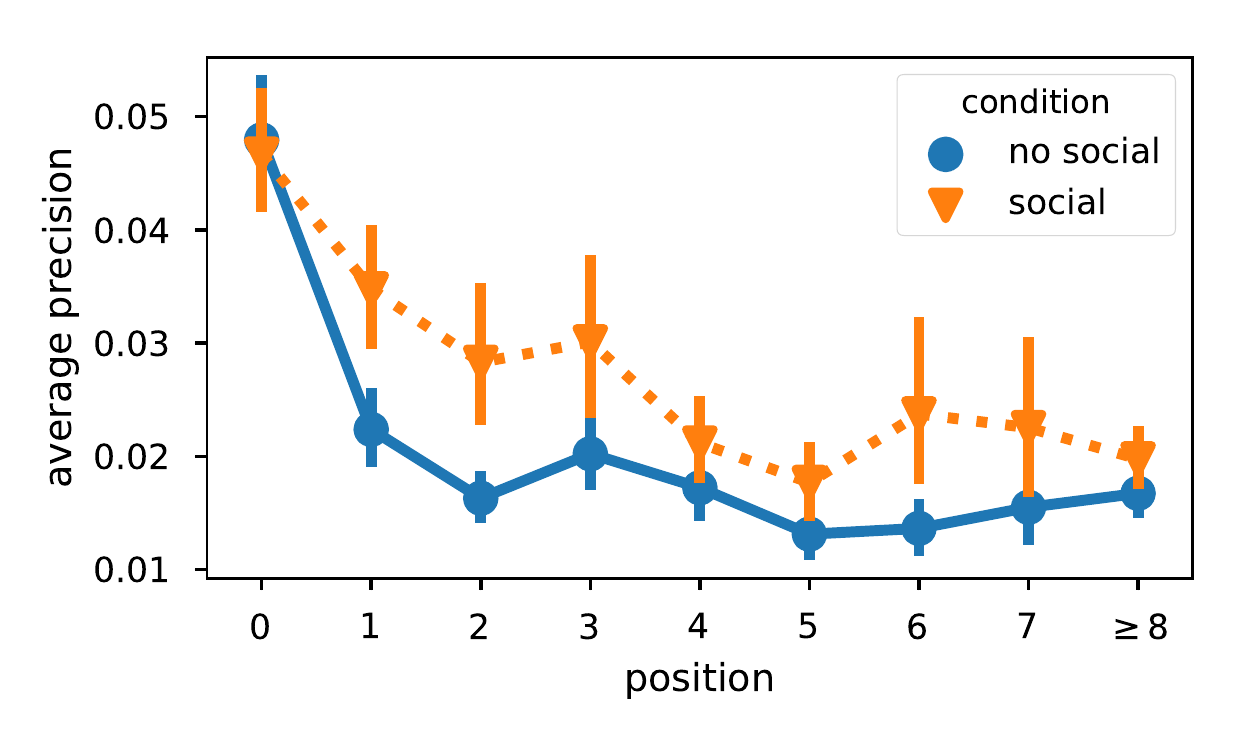}
  \vspace*{-0.3cm}
  \caption{Ranking performance for the legislation co-sponsorship cascades, organized by the position of each co-sponsor in the cascade. Position $0$ is the initial sponsor.}
  \label{fig:app-leg}
\end{figure}

The campaign finance network significantly improved ranking performance, $p \ll .0001$ by a paired $t$-test on the mean average precisions across cascades ($t=7.9$). The social influence features improve ranking performance on out-of-sample data as shown in \autoref{fig:app-leg}. These improvements are particularly strong for the earliest co-sponsors, who may be more likely to be motivated by campaign finance considerations. 

\subsection{Scientific rumors}
\label{sec:scientific-rumor}
On 4th July, 2012, the Higgs boson particle was discovered. 
Before this discovery was announced, rumors began on social media that the particle might indeed be the Higgs boson.
These ``scientific rumors'' on Twitter were collected by \cite{de2013anatomy} to study dynamics behind their spread.
We applied our discriminative ranker to predict the users most likely to spread the rumor.

\subsubsection{Data}
\citeauthor{de2013anatomy} collected tweets from users who mentioned any phrase from a set they identified to be about the Higgs boson discovery from July 1 to July 7, 2012. In conjunction, they collected the friends-follower, retweet, reply and mention networks between users of these tweets.
Some network statistics about this data\footnote{Available at http://snap.stanford.edu/data/higgs-twitter.html} is given in Table~\ref{table:network-statistics}.
For our work, we used the strongly-connected giant component of the friends-follower network and the cascade of first 5000 retweets spanning approximately the first two days of the cascade.

\subsubsection{Problem setup and evaluation}
To predict the users who spread the rumor, we rank them by training the discriminative ranker on all but the last 100 events of the cascade using all features described in Table~\ref{table:features}. 
The accuracy of ranking is then computed on these remaining 100 events, and is evaluated with average precision. 

\subsubsection{Baselines}
We set up two baselines. To compare the quality of predictions with chance, in the first baseline users are ranked randomly. In the second baseline, users are ranked by their number of past events (tweets about the Higgs boson); for users with no such events, we rank by network degree, with the intuition that this is a rough proxy for their rate of tweeting.

\subsubsection{Results}
\begin{table}
\caption{Average precision of predicting the next users to spread rumors of the Higgs boson discovery.}
  \label{table:higgs-results}
  \centering
  \begin{tabular}{ll}
    \toprule
    Rankers & Average precision\\
    \midrule
    Baseline 1 (random) & 0.0002\\
    Baseline 2 (by tweeting rate) & 0.0008 \\[1ex]
    Discriminative ranker & 0.0014 \\
  \bottomrule
\end{tabular}
\end{table}

Results are shown in \autoref{table:higgs-results}. As indicated by the baseline performance, this is a very difficult ranking task, since any of the $4.5 \times 10^5$ nodes in the network could potentially be the next to share the rumor. Ranking the nodes by their tweeting rate improves the average prediction somewhat, but the discriminative ranker yields the best overall performance. This demonstrates the importance of combining activity rate and network structure in this task.

%% file: discussion.tex
\section{Discussion}
\label{sec:discussion}
The experiments on both synthetic and real data demonstrate the utility of discriminative ranking for the detection of social influence and
the prediction of cascades.  
Despite this evidence in favor of the ranking test, it has limitations.
First, Granger causality is a limited, \emph{predictive} notion of causality; the detection of Granger causality in an event cascade does not necessarily imply that an intervention into the cascade at time $t$ would affect subsequent events. More concretely, it is generally not possible to distinguish social influence from homophily using observational data without taking the assumptions that all homophily-related confounds are observed and properly specified~\citep{shalizi2011homophily}.
Our approach is built on the assumption that homophily is effectively proxied by social network node embeddings.
While such assumptions are required of any test that is based on observational data, the ranking test still has advantages over other such tests. 
The discriminative nature of the ranker allows easy addition of a variety of features to improve prediction, including social network node embeddings that can summarize structural information that is likely to be related to unobserved confounds~\citep[see also contemporaneous work by][]{veitch2019using}. Researchers such as \cite{hofman2017prediction} have argued that social scientific methods should have predictive strength in addition to explanatory power, and we view this discriminative ranking test as an example of that vision.

More technically, the ranker comparison is based on a permutation test, which is a non-parametric distribution-free approach.
When the predictions of the two rankers are very similar, the resulting distribution can have low variance, leading to conservative p-values. 
This problem is especially severe in three cases: when the network size is small; when the heldout set is small; and when the base ranker is already very accurate, leaving little room for improvement with the addition of the social features.
The conservativeness of the permutation test does not affect the validity of the test~\citep{edgington2007randomization}, but it can affect the power.
Future work may explore alternative methods for comparing rankers.

There are also several avenues for future research.
We motivated the synthetic data experiments by practical constraints on the data, but more complex scenarios can occur in real world and their impact on our test needs to be empirically tested.
For example, missing data can itself be correlated with the social network, as in the case that neighboring nodes in a social media network decide to make their posts private.
Another direction for future work is to adapt our discriminative ranker to dynamic networks.
In our discriminative approach, this should be possible by using node embeddings at the input layer which can then be updated during learning as the network changes over time.


%% file: acknowledgments.tex
\section*{Acknowledgments}
We thank the reviewers for their constructive feedback that helped improve the paper. 
We also thank all members of the computational linguistics lab at Georgia Tech, who provided feedback on early drafts. 
The work also benefited from discussion at various times with Mart\'{o}n Karsai, Chris Naiqing Gu and Benjamin Carterette.
This research was supported by NSF award 1452443 and NIH award R01-GM112697-03.